\documentclass[11pt]{article}
\usepackage{moriond,epsfig}

\def\Journal#1#2#3#4{{#1} {\bf #2}, #3 (#4)}

\def\PLB{{\em Phys. Lett.}  B}
\def\PRL{\em Phys. Rev. Lett.}

\def\Phrep{{\em Phys. Rep.}}

\def\be{\begin{equation}}
\def\ee{\end{equation}}
\def\bea{\begin{eqnarray}}
\def\eea{\end{eqnarray}}

\newcommand{\Met}{\mbox{$E\!\!\!\//_{T}$}}
\newcommand{\stopa}{${\tilde{t}_{1}}$}
\newcommand{\sbota}{${\tilde{b}_{1}}$}

\newcommand{\sneuttau}{$\tilde{\nu}_{\tau}$}
\newcommand{\astopa}{${\bar{\tilde{t}}_{1}}$}

\newcommand{\fb}{$\rm fb^{-1}$}

\newcommand{\msta}{\mbox{$m_{\tilde{t}_1}$}}
\newcommand{\msneut}{$m_{\tilde{\nu}}$}

\newcommand{\chara}{$ \tilde{\chi}_{1}^{\pm} $ }
\newcommand{\mchara}{$ m_{\tilde{\chi}_{1}^{\pm}} $ }

\newcommand{\pt}{$p_{T}$}

\newcommand{\deltam}{$\Delta m$ }
\newcommand{\deltamdef}{$\Delta m=m_{\tilde{t}_1}-m_{\tilde{\nu}}$ }

\newcommand{\bchar}{$b \tilde{\chi}_{1}^{+}$ }

\newcommand{\blepsneut}{$b l \tilde{\nu}$}

\newcommand{\dzero}{D\O}

\newcommand{\neuta}{$\tilde{\chi}_1^0$}
\newcommand{\mneuta}{$m_{\tilde{\chi}_1^0}$}
\newcommand{\neutb}{$\tilde{\chi}_2^0$}

\newcommand{\msbottom}{$m_{\tilde{b}_{1}}$}

\newcommand{\sneut}{$\tilde{\nu}$}

\newcommand{\ttbar}{$t\bar{t}$}
\newcommand{\ddbar}{$d\bar{d}$}

\newcommand{\ztautau}{\mbox{$Z/\gamma^*\rightarrow \tau^+\tau^-$}}

\newcommand{\darkg}{$\gamma_D$}

\newcommand{\gravit}{$\tilde{G}$}

\begin{document}
\vspace*{1.3cm}
\title{SUSY Searches at the Tevatron}

\author{ Ph. Gris \\
(for the CDF and \dzero~collaborations)}

\address{Laboratoire de Physique Corpusculaire,\\
IN2P3/CNRS, 24 avenue des Landais , 63177 Aubi\`ere cedex, France}

\maketitle\abstracts{
The results of search for Supersymmetry performed at the Tevatron Collider by the CDF and \dzero~collaborations are summarized in this paper. No significant deviations with respect to the Standard Model expectations were observed and constraints were set on supersymmetric parameters.}

\section{Introduction}
Supersymmetry~\cite{susy} (SUSY), a space-time symmetry that predicts for every Standard Model (SM) particle the existence of a superpartner that differs by half a unit of spin, may provide a solution to the hierarchy problem if SUSY particles have masses lower than 1 TeV, strongly motivating the search for such particles at the Fermilab Tevatron Collider. If there is supersymmetry in nature, it must be broken and the theorized breaking mechanisms lead to many models (supergravity, gauge mediated, anomaly mediated, ...) with possibly different phenomenologies. Searches performed by the CDF and \dzero~experiments aim at probing the extensive SUSY parameter space in terms of mass and final state.

\section{Third generation squark searches}

Because of the large Yukawa couplings of the third-generation quarks, the lightest stop \stopa~and sbottom \sbota, partners of the top and bottom quarks respectively, may be the lightest squarks and have masses reachable at the Tevatron.

\subsection{Stop searches}

\subsubsection{\stopa $\rightarrow$ \blepsneut}
A search for stop pairs has been performed by the \dzero~collaboration~\cite{dzero_stop} in a dataset corresponding to an integrated luminosity of 5.4 \fb. The three-body decay chosen for the stop, \blepsneut, lead to a final state containing two leptons ($e$ and $\mu$), two {\it b}-jets and missing transverse energy (\Met). The sneutrino $\tilde \nu$ is either the Lightest Supersymmetric Particle (LSP) or decays invisibly. The signal topology depends on the difference of the stop and sneutrino masses \deltamdef: the transverse momentum (\pt) of the leptons and jets tend to decrease with \deltam. The analysis was thus optimised for two mass domains (\deltam~above or below 60 GeV). The main background is composed of Drell-Yan events \ztautau. No significant deviation from SM predictions has been found. The results were translated into 95\% C.L. limits in the (\msta,\msneut) plane \mbox{(Fig. \ref{fig:figa}, left)}. 

\subsubsection{\stopa $\rightarrow$ \bchar $\rightarrow$ $b\ell\nu$\neuta}
\begin{figure}
\begin{center}
\epsfig{figure=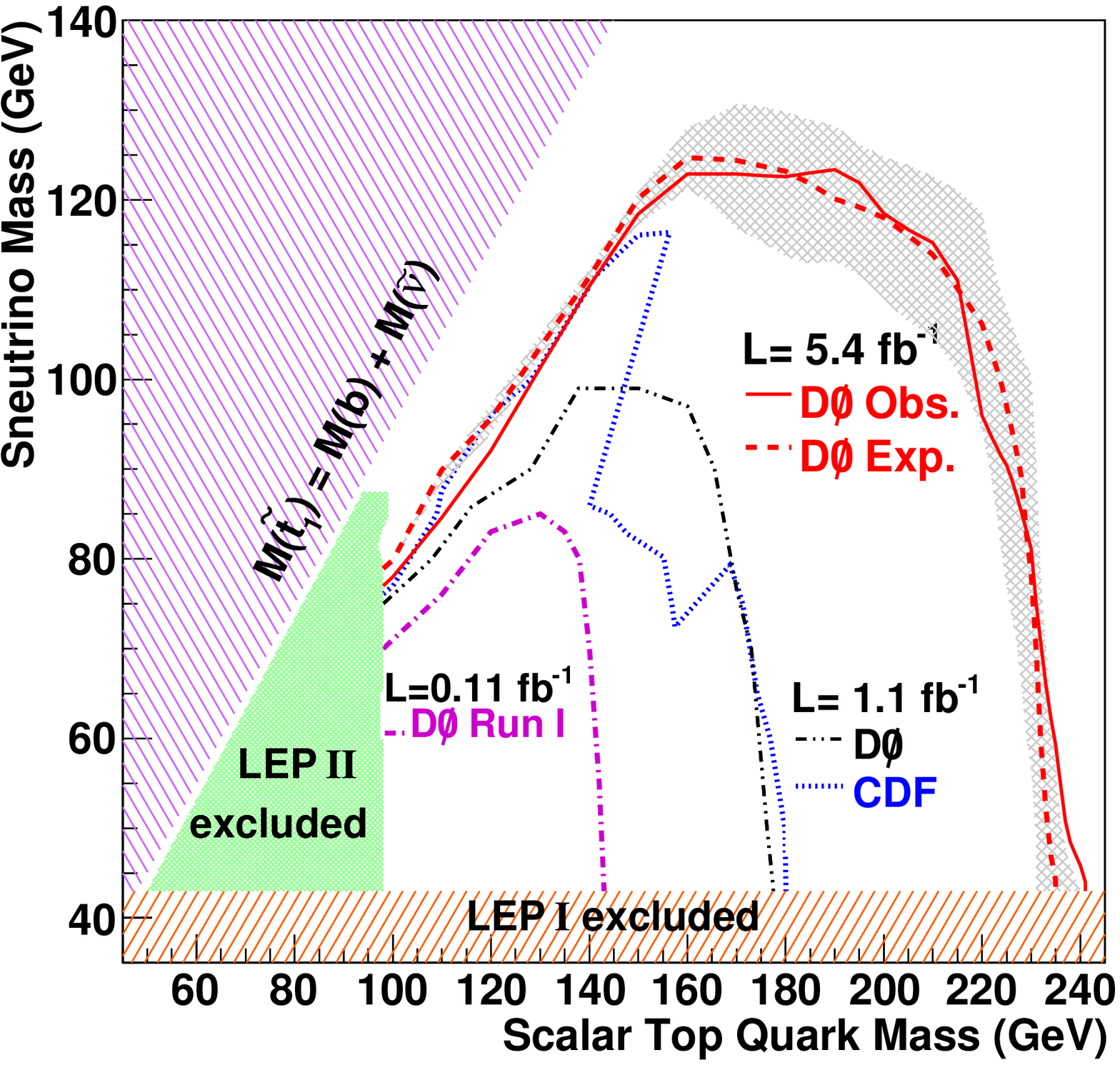,height=4.5cm,width=5.5cm}
\epsfig{figure=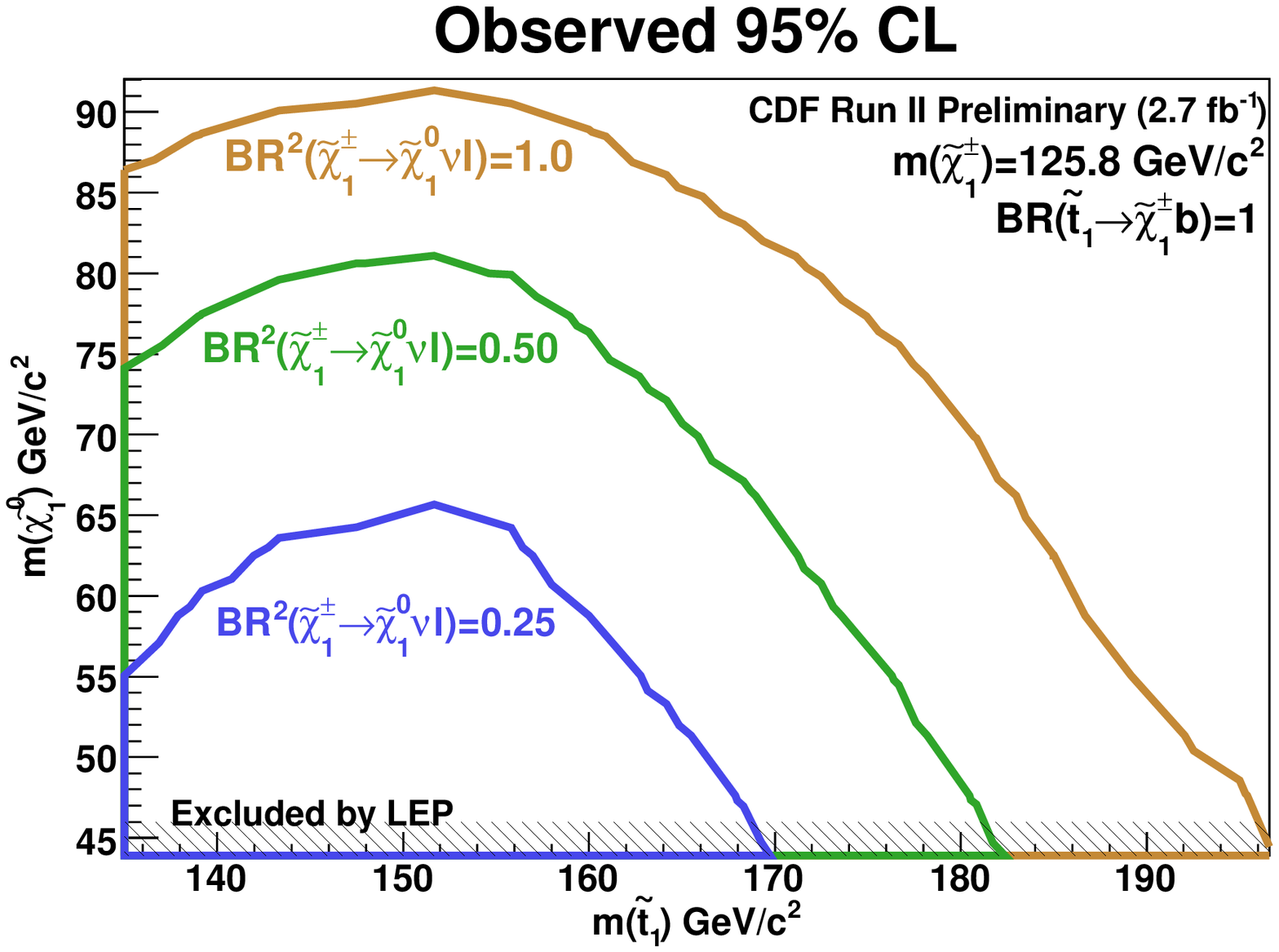,height=4.5cm,width=5.5cm}
\end{center}
\caption{95\% C.L. limits extracted from stop searches. The three-body decay \blepsneut~was considered by the \dzero~experiment (left) whereas  the CDF collaboration has studied \stopa$\rightarrow$\bchar (right).
\label{fig:figa}}
\end{figure}

A search for the two-body decay of the stop \stopa $\rightarrow$ \bchar, has been performed by the CDF experiment~\cite{cdf_stop}. In the scenario considered, the lightest neutralino \neuta~is the LSP and the chargino decays to $b\ell\nu$\neuta. The final state is composed of two leptons ($ee$,$e\mu$,$\mu\mu$), two {\it b}-jets and \Met. The selection criteria used in the analysis aim at removing the SM backgrounds ($W/Z$+jets, diboson, \ttbar, QCD) while keeping a high efficiency for the signal: the presence of at least two leptons and two jets is requested, as well as minimal \Met~(20 GeV). Classes of events were created depending on their heavy flavour content. The mass of the top squark candidate events were reconstructed and have provided discrimination between a \stopa\astopa~signal and SM backgrounds. In a sample corresponding to an integrated luminosity of 2.7 \fb, data were found to be consistent with the expectations from SM processes and the results were used to extract the 95\% C.L. exclusion limit in the \mneuta~vs \msta~plane for several values of the branching ratio $BR($\chara$\rightarrow$$\ell \nu$\neuta$)$ and \mbox{\mchara,} assuming equal branching ratio into different lepton flavour and $BR($\stopa$\rightarrow$ \bchar$)$=100\% \mbox{(Fig.~\ref{fig:figa}, right)}. 

\subsection{Sbottom searches}

Assuming a SUSY spectrum mass hierarchy such that the sbottom decays exclusively as \sbota$\rightarrow$$b$\neuta, the expected signal for direct sbottom pair production is composed of two {\it b}-jets and \Met~coming from the two \neuta-LSP in the final state. The \dzero~collaboration~\cite{dzero_sbottom} has searched for direct sbottom production in a data sample corresponding to an integrated luminosity of 5.2 \fb. Events were selected if they contained at least two jets, no lepton and a significant \Met~(at least 40 GeV). Events with {\it b}-jets in the final state were retained using a Neural Network (NN) b-tagging algorithm. The dominant source of background was composed of events with a light flavour jet mis-identified as a {\it b}-jet (mistag) and was estimated from data, whereas other backgrounds (except QCD) were evaluated using Monte Carlo (MC) simulations. No significant deviation from the SM background prediction has been observed and the results have been interpreted as a 95\% C.L. exclusion limit in the \mneuta~vs \msbottom plane.

\section{Gaugino searches}

If squarks and gluinos are too heavy to be produced at rates high enough to be currently detected at the Tevatron, a search for SUSY can be performed via the associated production of charginos and neutralinos \chara\neutb.

\subsection{Search for trilepton events}

In the scenario considered the gauginos decay via exchanges of vector bosons or sleptons into \neuta(LSP) and SM fermions: \chara$\rightarrow$$\ell\nu$\neuta, \neutb$\rightarrow$$\ell\ell$\neuta. The CDF~\cite{cdf_trilept} and \dzero~\cite{dzero_trilept} experiments have searched for final states containing three leptons and \Met~in data samples corresponding to integrated luminosities of 3.2 and 2.3 \fb, respectively. Classes of events were selected depending on the quality (loose or tight) and the number of isolated leptons ($e,\mu,\tau$). Events with two identified leptons plus an isolated track were also retained. The background composed of real leptons (from diboson, \ttbar, $Z$+jets processes) was estimated from MC simulations whereas fake leptons(jets faking electrons, tracks faking muons) were extracted from data. After selection the observed data was found to be consistent with SM predictions and 95\% C.L. exclusion limits on the production cross section and leptonic branching fraction have been estimated from data in the framework of the minimal SuperGRAvity (mSUGRA) model.

\subsection{Search for diphoton + MET events}

In Gauge Mediated Supersymmetry Mediated (GMSB) models, the superpartner of the graviton, the gravitino \gravit, may be the LSP and have a very low mass (few keV). The lightest neutralino \neuta~may be the Next-LSP (NLSP) and decays through: \neuta$\rightarrow$$\gamma$\gravit. \neuta~may be produced in cascade decay chains coming from the production of \chara\chara~or \chara\neutb. The \dzero~experiment~\cite{dzero_gmsb} has searched for events with at least two photons and large \Met (at least 35 GeV) in a data sample corresponding to an integrated luminosity of 6.3 \fb. No explicit requirement for additional presence of jets and leptons has been imposed. The dominant SM background arising from instrumental \Met~sources (SM $\gamma\gamma$, $\gamma$+jets, multijet) and from genuine \Met ($W\gamma$,$W$+jets) was estimated from control data samples. No evidence for SUSY signal has been found and limits at the 95\% C.L. were extracted for a specific set of the GMSB parameters that correspond to the Snowmass Slope constraints SPS8. A chargino (neutralino) mass lower than 330 GeV (175 GeV) has been excluded.
 
\subsection{Search for leptonic jets+MET}

Hidden valley models~\cite{hidden_model} introduce a new hidden sector weakly coupled to the SM. A large subset of hidden valley models also contain supersymmetry. In this scenario, the force carriers in the hidden sector are the dark photons \darkg. The lightest neutralino \neuta~is not the LSP and can further decay into a hidden sector dark neutralino plus a photon or a dark photon. \darkg~ultimately decays into a pair of spatially closed leptons (the leptonic jet) if m(\darkg)$\leq$ 2m$(\pi)$, mass range of interest to explain astrophysical anomalies. The \dzero~experiment~\cite{dzero_hidden}  has performed a search of the production \chara\neutb~in a 5.8 \fb~data sample. Events were selected by requiring two leptonic jets ($\mu$ or $e$) in each event, the three classes $\mu\mu$, $e\mu$ and $ee$ being treated separately. The main background arising from multijet production and photon conversion was estimated from data. 21 events survived the selections whereas 36.3$\pm$10.4 were expected. This result was translated as a 95\% C.L. limit on the production cross section of the \chara\neutb~process.

\section{$R_{parity}$ violating searches}

In SUSY theories, $R_{parity}$\cite{rpv} is a quantum number equal to +1 for SM particles and -1 for superpartners. For searches presented above, $R_{parity}$ was assumed to be conserved and LSP was stable. If it is not the case, LSP can decay into SM particles.

\subsection{Search for tau sneutrino}

\begin{figure}
\begin{center}
\epsfig{figure=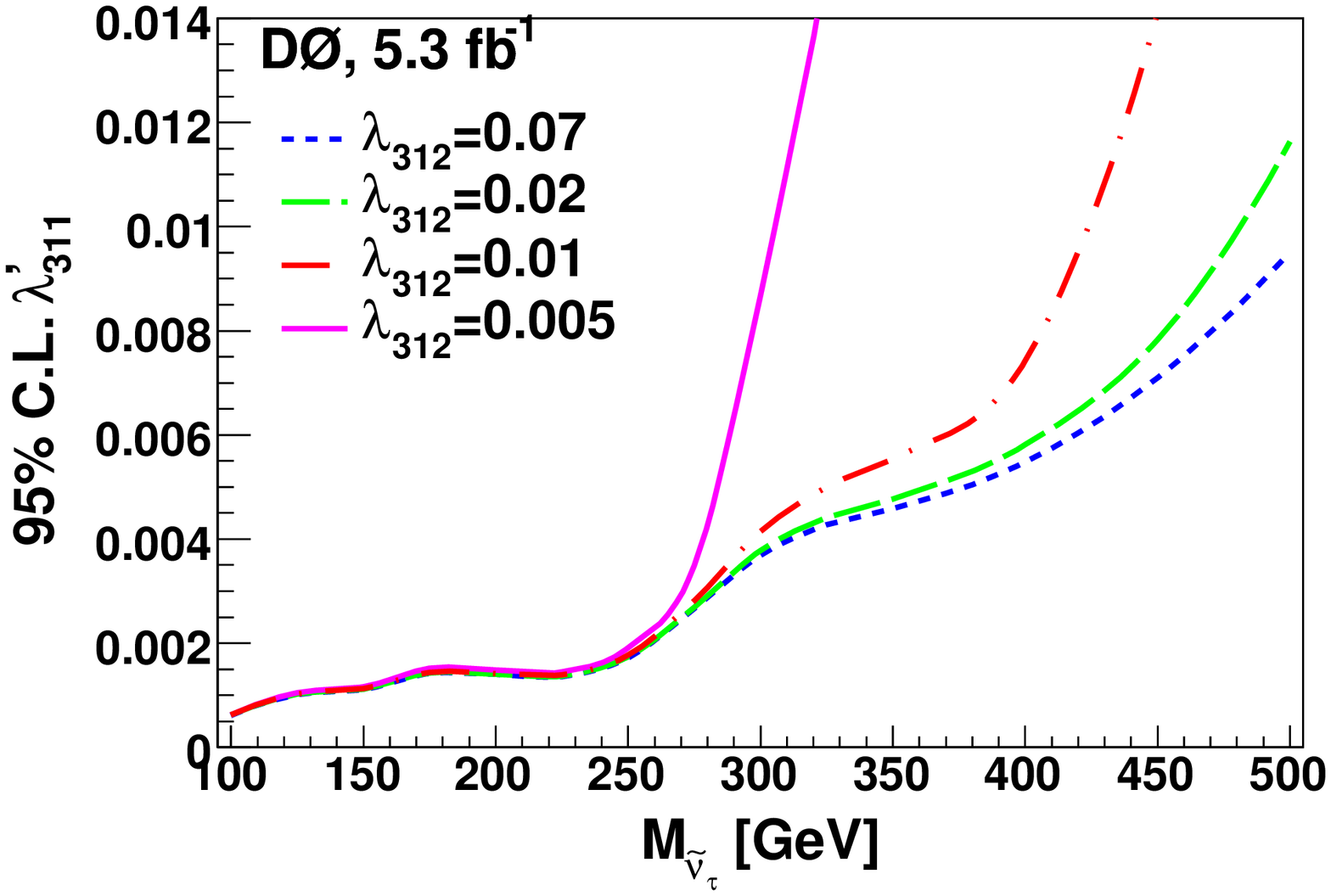,height=4.5cm,width=6.cm}
\epsfig{figure=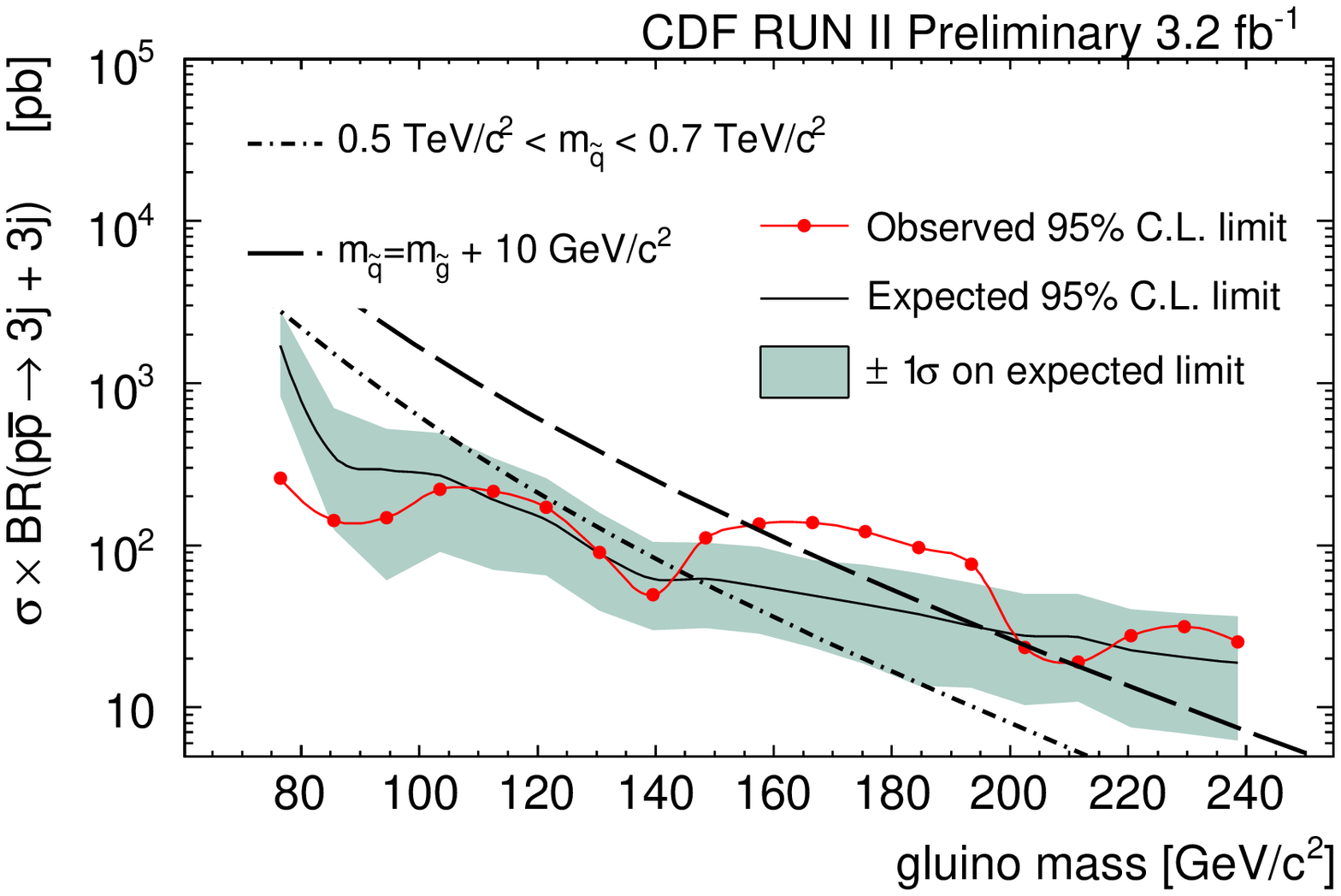,height=5.cm,width=6.5cm}
\end{center}
\caption{\dzero~cross section limit on the resonant sneutrino as a function of the sneutrino mass (left) and CDF cross section limit on the gluino pair production as a function of the gluino mass (right).
\label{fig:figb}}
\end{figure}

If $R_{parity}$ is violated the sneutrino \sneut~may be produced in \ddbar~scattering and thus produced at the Tevatron. The \dzero~collaboration~\cite{dzero_sneut} has searched for the resonant production of the tau sneutrino \sneuttau~($\lambda_{311}^{'}$ coupling) decaying to $e\mu$ ($\lambda_{321}$ coupling). Events were selected by requiring two isolated leptons ($E_T \geq $ 30 GeV) and no jet. 414 events survived the final cuts whereas 410$\pm$38 were expected from MC simulations. No evidence for tau sneutrino signal has been found in the 5.3 \fb~data sample analysed. The resulting 95\% C.L. limit on the cross section led to exclude sneutrino masses heavier than 300 GeV (Fig. \ref{fig:figb}, left).

\subsection{Search for 3-jet resonances}

The CDF experiment~\cite{cdf_threejets}  has performed a model independent search for 3-jet hadronic resonances. Events with a high number of jets (at least six) and no \Met~were selected. The dominant multijet background was estimated from data using 5-jet events. To model new physics signatures that could hide in the analyzed data sample corresponding to an integrated luminosity of 3.2 \fb, gluino pairs decaying each into three partons (through $\lambda^{''}$ couplings) were chosen. Kinematic quantities such as invariant mass and scalar \pt~sum of jet triplets (and their correlations) were used to extract the signal. No significant excess of data has been observed and 95\% C.L. limits on the production cross section of the process \mbox{$\tilde{g}\tilde{g}\rightarrow$3 jets+3 jets} have been set: gluino masses below 144 GeV have been excluded (Fig. \ref{fig:figb}, left).

\section{Summary}
This contribution reported the most recent results of searches for SUSY at the Tevatron. Analyses performed by the CDF and \dzero~collaboration using up to 6.3 \fb~of data were summarized. No significant data excess has been observed and the results have been interpreted as 95\% C.L. exclusion limits on the free parameters of several SUSY scenarios.


\section*{References}
\begin{thebibliography}{99}

\bibitem{susy} See, for instance:
               H.P. Nilles, \Journal{\Phrep}{110}{1}{1984}. 

\bibitem{dzero_stop} V.M. Abazov {\it et al}, \Journal{\PLB}{696}{321}{2011}.

\bibitem{cdf_stop} T. Aaltonen {\it et al}, \Journal{\PRL}{104}{251801}{2010}.

\bibitem{dzero_sbottom} V.M. Abazov {\it et al}, \Journal{\PLB}{693}{95}{2010}.

\bibitem{cdf_trilept}  CDF public note 9817.

\bibitem{dzero_trilept} V.M. Abazov {\it et al}, \Journal{\PLB}{680}{34}{2009}.

\bibitem{dzero_gmsb} V.M. Abazov {\it et al}, \Journal{\PRL}{105}{221802}{2010}.

\bibitem{hidden_model} M. Strassler and K. Zurek, \Journal{\PLB}{651}{374}{2007}.

\bibitem{dzero_hidden} V.M. Abazov {\it et al}, \Journal{\PRL}{105}{211802}{2010}.

\bibitem{rpv} For a review see: R. Barbier {\it et al}, \Journal{\Phrep}{420}{1}{2005}. 

\bibitem{dzero_sneut} V.M. Abazov {\it et al}, \Journal{\PRL}{105}{191802}{2010}.

\bibitem{cdf_threejets} CDF public note 10256.

\end{thebibliography}
\end{document}